\documentclass[aps,twocolumn,showpacs,preprintnumbers,amsmath,amssymb]{revtex4}
\usepackage{graphicx}% Include figure files
\usepackage{dcolumn}% Align table columns on decimal point
\usepackage{bm}% bold math

\newcommand{\co}[2]{\ifcase #1 \or #2 \fi}
\newcommand{\lcco}{{La$_{2-x}$Ce$_{x}$CuO$_4$}}
\newcommand{\ncco}{{Nd$_{2-x}$Ce$_{x}$CuO$_4$}}
\newcommand{\pcco}{{Pr$_{2-x}$Ce$_{x}$CuO$_4$}}

\begin{document}

\preprint{Wagenknecht et al./ZBCA-Hc2 -- vers.01}

\title{Andreev bound states at a cuprate grain boundary junction: A lower bound for the upper critical field}
% Force line breaks with \\

\author{M.~Wagenknecht}
\email{wagenknecht@uni-tuebingen.de}
\author{D.~Koelle}
\author{R.~Kleiner}
\affiliation{
Physikalisches Institut --
Experimentalphysik II,
Universit\"{a}t T\"{u}bingen,
Auf der Morgenstelle 14,
D-72076 T\"{u}bingen,
Germany
}
\author{S.~Graser}
\author{N.~Schopohl}
\affiliation{
Institut f\"{u}r Theoretische Physik,
Universit\"{a}t T\"{u}bingen,
Auf der Morgenstelle 14,
D-72076 T\"{u}bingen,
Germany
}
\author{B.~Chesca}
\affiliation{
Department of Physics,
Loughborough University,
Loughborough,
Leics LE11 3TU,
United Kingdom
}
\author{A.~Tsukada}
\affiliation{
NTT Basic Research Laboratories, 3-1 Morinosato Wakamiya, Atsugi-shi, Kanagawa 243, Japan}%

\author{S.~T.~B.~Goennenwein}
\author{R.~Gross}
\affiliation{
Walther-Meissner-Institut, Bayerische Akademie der Wissenschaften, Walther-Meissner Str. 8, D-85748 Garching, Germany}%

\date{\today}

\begin{abstract}
We investigate in-plane quasiparticle tunneling across thin film
grain boundary junctions (GBJs) of the electron-doped cuprate \lcco,
in magnetic fields up to $B=16\,$T, perpendicular to the CuO$_2$
layers.
The differential conductance in the superconducting state shows a
zero bias conductance peak (ZBCP) due to zero energy surface Andreev
bound states.
With increasing temperature $T$, the ZBCP vanishes at the critical
temperature $T_c\approx29\,$K if $B=0$, and at $T=12\,$K for
$B=16$\,T.
As the ZBCP is related to the macroscopic phase coherence of the
superconducting state, we argue that the disappearance of the ZBCP at
a field $B_{ZBCP}(T)$ must occur below the upper critical field
$B_{c2}(T)$ of the superconductor.
We find $B_{ZBCP}(0) \approx 25\,$T which is at least a factor of 2.5 higher
than previous estimates of $B_{c2}(0)$.
\end{abstract}

 % PACS, the Physics and Astronomy Classification Scheme.
                                              
\maketitle
%Introduction
Determining the magnetic field$-$temperature ($B-T$) phase diagram of high-$T_c$ cuprates has been the focus of interest since the discovery of these materials. In contrast to the case of conventional type II superconductors, where the $B-T$ phase diagram basically consists of the Meissner phase, the Shubnikov phase and the normal state, the phase diagram of high-$T_c$ cuprates is extremely rich, exhibiting a variety of vortex phases and also a pseudogap region. However, particularly the transition between the superconducting state and the normal state, and thus the relation between the superconducting and the pseudogap states, is hard to determine, not only due to the extremely large values of the upper critical field $B_{c2}$ in the case of hole-doped cuprates, but also because of the presence of vortex liquid phases as well as strong fluctuation effects, leading to nonzero resistance in transport experiments long before $B_{c2}$ has been reached. 

In order to evaluate similarities and differences to their hole-doped counterparts, many investigations have been performed within the past years on electron doped materials like the single layer T' cuprates \cite{Tokura89} with composition $Ln_{2-x}$Ce$_x$CuO$_4$ ($Ln$ = La, Pr, Nd). \pcco (PCCO) and \ncco (NCCO) have been studied extensively,
while there is less data published for \lcco (LCCO).
The precise determination of the phase diagram of these materials is certainly of fundamental interest.

In terms of $B_{c2}$, for PCCO and NCCO thin films near optimal doping, resistive measurements \cite{Fournier03} or an analysis of the vortex Nernst signal \cite{Gollnik98, Balci03, Wang03} revealed $B_{c2}(0)$ values in the range of 7-10$\,$T.
For optimally doped LCCO an analysis of the vortex pinning strength yielded $B_{c2}(0)$ around 9$\,$T \cite{Zuev03}. However, the various methods applied to determine the upper critical field often yield inconsistent results, see e.~g.~the discussion in \cite{Fournier03}. 
In this paper we show that an analysis of Andreev bound states (ABS) causing a zero bias conductance peak (ZBCP) in the conductance spectra of cuprate grain boundary junctions (GBJs) yields a new lower bound for $B_{c2}$ which is at least a factor of 2.5 above previous estimates.

ABS result from the constructive interference of Andreev reflected electron like and hole like quasiparticles. If the quasiparticles experience a sign change of the superconducting order parameter upon reflection, the ABS appear at the Fermi energy, giving rise to the ZBCP \cite{Kashiwaya00}. ABS-caused ZBCPs have been observed both in hole-doped
\cite{Ekin97, Covington97, Alff97, Alff98, Wei98} and electron-doped \cite{Hayashi98, Mourachkine00, Biswas02, Chesca05}  cuprates, where the sign change is due to the $d_{x^2-y^2}$ symmetry of the order parameter \cite{Tsuei00RMP}. This type of ZBCP can be observed with GBJs \cite{Alff98, Chesca05} as well as for normal metal/superconductor junctions \cite{Ekin97, Covington97, Alff98, Wei98, Hayashi98, Mourachkine00, Biswas02} or junctions between a conventional superconductor and a $d$-wave superconductor \cite{Chesca06}.

We note here that ZBCPs in the quasiparticle tunnel spectrum of superconductors can have other reasons than the formation of ABS, most prominent being the scattering of quasiparticles at magnetic impurities situated in the tunnel barrier, as described
by the Appelbaum-Anderson model \cite{Appelbaum66, Anderson66}. Such competing mechanisms can  be identified, however, e.~g.~by analyzing  the temperature and magnetic field depencence of the ZBCP.

A ZBCP caused by ABS relies on the phase coherence of the elementary excitations above the Cooper pairing ground state.
It thus should vanish when the macroscopic phase coherence is lost, i.~e., at the transition between the superconducting state and the normal state.
Such ZBCPs thus allow, at least in principle, to determine $B_{c2}$ or at least to give a reasonable lower bound, provided that the ZBCPs do not vanish far below $B_{c2}$, e.~g.~due to the formation of a vortex liquid or other effects like a (field induced) shift of ABS to larger energy. 
As we will see below, in the geometry used in our experiments
an unsplitted ZBCP is clearly visible above the irreversibility line, allowing its use as a probe of the superconducting state in high magnetic fields. 

The 900$\,$nm thin films for our study were deposited by molecular beam epitaxy with ozone as the oxidation gas \cite{Naito00} on SrTiO$_3$ (STO) symmetric ($\alpha_1=-\alpha_2$, see inset of Fig.~\ref{fig:2}) [001]-tilt bicrystal substrates with a total misorientation angle ($\alpha_1+\alpha_2$) of 24$^\circ$ and 30$^\circ$. 
All films in this study had a Ce doping of $x\sim0.08$ (slightly underdoped, $T_c\approx29\,$K).
The samples were patterned by photolithography and Ar ion milling with junction widths ranging from 40$\,\mu$m
to 400$\,\mu$m 
\footnote{In fact we used a SQUID design to investigate both Cooper pair tunneling at low magnetic fields ($\mu$ T range) and quasiparticle tunneling at high fields (T range). For the measurements presented here Cooper pair tunneling plays no role.}.
Transport data were collected with a current bias four point probe at 5$\,$K$ \leq T \leq 40\,$K and magnetic fields $B \leq 16\,$T applied parallel to the thin film c-axis.
$dI(V)/dV$-characteristics were measured with a lock-in amplifier (modulation frequency 1-5$\,$kHz,
$I_{ac}=1-10\,\mu$A) and subsequently calibrated with the numerical derivative of simultaneously recorded $I(V)$-curves. Unless indicated otherwise, the data shown below are all for the same 24$^\circ$ GBJ.  

%Figure 1
\begin{figure}[tb]
\begin{center}
\includegraphics[width=0.9\columnwidth,clip]{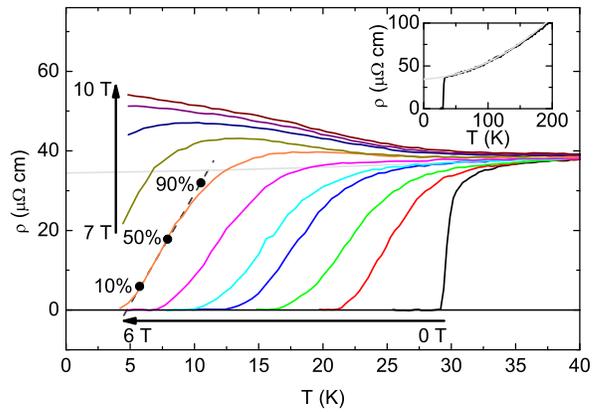}
\end{center}
\caption{(color online) $\rho(T)$ of a LCCO thin film for $0\,$T $\leq B \leq 10\,$T (in steps of 1$\,$T).
Dashed lines and full circles on the 6$\,$T curve indicate determination of temperatures where the resistance has dropped to,
respectively 10, 50 and 90 per cent of its normal state value (grey line). These values define the critical fields $B_{10}$, $B_{50}$ and $B_{90}$.
The inset shows $\rho(T)$ at $B = 0$
up to 200$\,$K with a parabolic fit (grey line).
\label{fig:1}} 
\end{figure}
%Figure 1 \end
Fig. \ref{fig:1} shows resistivity $\rho(T)$ of a bridge structure not crossing the grain boundary, for $0\,$T $\leq B \leq 10\,$T. Similar as for thin films of PCCO \cite{Kleefisch01} and LCCO \cite{Sawa02}, the onset of $T_c$ decreases with increasing $B$ and the transition width is broadened. For $B > 6\,$T the resistance remains nonzero for $T$ down to 5$\,$K, and for larger fields we also observe an upturn in $\rho(T)$, as already reported previously \cite{Fournier98, Kleefisch01}. We determine "transition temperatures" where the resistance of the (extrapolated) linear part of $\rho (T)$ [cf. dashed line in Fig. \ref{fig:1}] has dropped to, respectively 10, 50 and 90 per cent of its normal state value, given by the intersection of the dashed line with the parabolic fit for $\rho(T)$ above $35\,$K. The corresponding "transition temperatures" or, respectively, transition fields $B_{10}$, $B_{50}$ and $B_{90}$, are shown further below in the $B-T$ diagram
of Fig.~\ref{fig:4}.
Although the resistive transition broadens significantly, these fields do not differ strongly at a given temperature and tend to extrapolate to a zero temperature value around 7$\,$T, suggesting that $B_{c2}(0)$ is of this order. However, as we will discuss below, ZBCPs can be seen in much higher fields, suggesting e.~g.~superconductivity below 12$\,$K at 16$\,$T. 

Fig. \ref{fig:2} (a) shows quasiparticle tunneling spectra (QTS) for a 24$^\circ$ GBJ, as obtained at 5$\,$K for $B$ up to 16$\,$T.
At $B = 0$, a clear gap structure, that is a suppression in the quasiparticle density of states (QDOS), is observed, with symmetric 
coherence peaks at voltages $V_{gap}=\pm 9\,$mV. 
There is a small dip feature of unknown origin at voltages near $\pm$ 7 mV which  disappears at higher fields. With increasing field the coherence peaks decrease and are strongly suppressed (although still present) for $B > 6\,$T.
For voltages well above $V_{gap}$ the conductance increases linearly with voltage, cf. upper inset in (b),  and is almost independent of the applied field up to 6$\,$T
\footnote{A similar linear increase of the background conductance at higher voltages also has been seen for other electron- or hole-doped cuprates, e.g. in \cite{Kleefisch01} and \cite{Dagan05}.}. 
For fields above 6$\,$T the differential conductance strongly decreases due to the onset of resistance in the film adding in series to the grain boundary resistance. This additional resistance has been subtracted in Fig. \ref{fig:2} (b), where one can see that the conductance in the subgap region increases monotonically with increasing field. The lower left  inset in Fig. \ref{fig:2} (b) shows the integral over the tunnel conductance, taken between -15$\,$mV and 15$\,$mV. The integral is field-independent, showing that the density of states is conserved.  

%Figure2
\begin{figure}[tb]
\begin{center}
\includegraphics[width=0.9\columnwidth,clip]{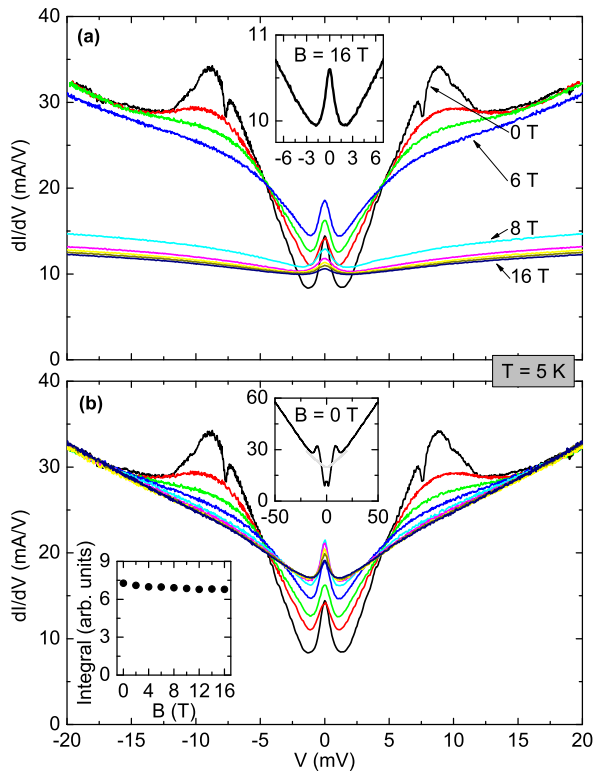}
\end{center}
\caption{(color online)
Quasiparticle conductance spectra of a 24$^\circ$ LCCO thin film GBJ at $T=5\,$K for $0\,$T $\leq B \leq 16\,$T in (steps of 2$\,$T).
Graph (a) shows conductance as measured. In (b) the resistance of the leads appearing above 6$\,$T has been subtracted.
The inset in (a) shows an enlargement of the low bias region at 16$\,$T.
The upper inset in (b) shows the zero field conductance on a larger voltage scale at $T=5\,$K (black line) and at $T=32\,$K (grey line),
where the ZBCP has disappeared. The lower left inset shows the integral over the conductance curves taken for voltages between -15$\,$mV and +15$\,$mV.
The lower right inset in (b) illustrates the geometry of the bicrystal GBJ. Vortices are indicated by circles, screening currents by arrows.
\label{fig:2}} 
\end{figure}
In the QTS of our samples we also observe a ZBCP, consistent with the previous study on optimally doped thin film LCCO GBJs \cite{Chesca05}. At $T=5\,$K the ZBCP persists up to the highest fields achievable with our setup, i.~e.~16$\,$T. The ZBCP can already be seen clearly in the uncorrected data, as shown in the inset of Fig. \ref{fig:2} (a). We saw similar ZBCPs in 15 out of 19 samples both on 24$^\circ$ and 30$^\circ$ substrates.

\begin{figure}[tb]
\begin{center}
\includegraphics[width=0.9\columnwidth,clip]{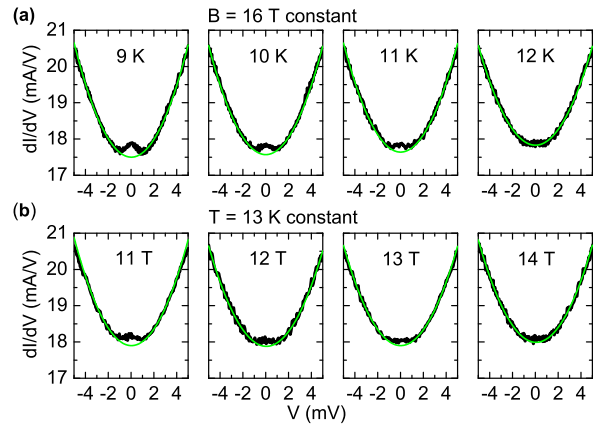}
\end{center}
\caption{(color online) Quasiparticle conductance $dI(V)/dV$ of a 24$^\circ$ LCCO thin film GBJ
in the vicinity of zero bias showing the disappearance of the ZBCP 
(a) at constant field $B\,=\,16\,$T for different $T$ and 
(b) at constant temperature $T\,=\,13\,$K for different $B$.
Thin lines are parabolic fits to the quasiparticle conductance outside the ZBCP. 
\label{fig:3}} 
\end{figure}

\begin{figure}[tb]
\begin{center}
\includegraphics[width=0.9\columnwidth,clip]{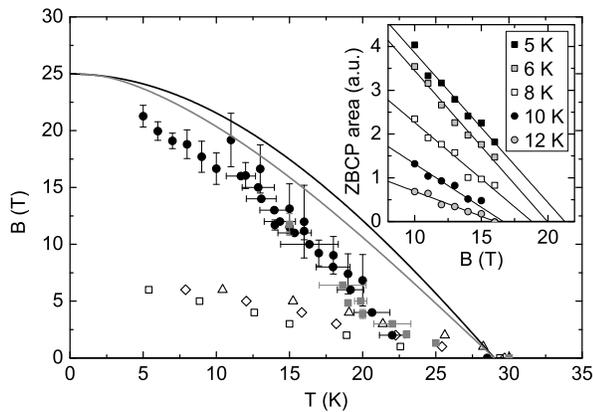}
\end{center}
\caption{(color online) Magnetic field vs. temperature phase diagram showing the fields $B_{10}$ (open squares), $B_{50}$ (open diamonds) and $B_{90}$ (open triangles), as determined from $\rho(T)$ curves of Fig. \ref{fig:1}, together with the field $B_{ZBCP}$ where the ZBCP in the quasiparticle tunnel spectrum  disappears. Full circles: 24$^\circ$ GBJ, full squares: 30$^\circ$ GBJ. The black and the grey lines correspond to $B_{c2}(T)$, calculated for a 2D Fermi cylinder and a 3D Fermi sphere, respectively. The inset shows the ZBCP area vs. $B$ for different temperatures. Lines are linear fits, which, by extrapolation to zero ZBCP area, define values of $B_{ZBCP}$ for $B > 16\,$T.
\label{fig:4}} 
\end{figure}

Fig. \ref{fig:3} shows the evolution of the ZBCP as a function of $T$ at $B = 16\,$T (a), and as a function of $B$ at $T = 13\,$K (b).
Data are shown by thick black lines. The thin lines represent parabolic fits to the background quasiparticle conductance at low voltages.
From the figures we see that, at $B = 16\,$T the ZBCP disappears between 11$\,$K and 12$\,$K, while at
$T = 13\,$K it disappears between 13$\,$T and 14$\,$T
\footnote{We note here that for fields where the ZBCP has disappeared the quasiparticle conductance looks parabolic but without a clear sign of an additional suppression at low voltages due to a pseudogap state.}. 

%Figure 3 \begin
%Figure 3 \end

As the ZBCP disappears reproducibly for a given $T$ or $B$ while
increasing $B$ or $T$, respectively, we can follow the disappearance
of the ZBCP in a $B - T$ phase diagram, which is shown in Fig.
\ref{fig:4}. Full symbols show the "critical" field $B_{ZBCP}$ where
the ZBCP area $A_{ZBCP}$ has reached zero for two different samples
(squares: 30$^\circ$ GBJ, circles:  24$^\circ$ GBJ). We determined
$A_{ZBCP}$ by integrating the difference of the measured
quasiparticle spectra and the parabolic background conductance
[c.~f.~Fig. \ref{fig:3}] in the range $\pm$ 3 mV. As
$A_{ZBCP}(B)_{|T=const}$ and $A_{ZBCP}(T)_{|B=const}$ decreases
nearly linearly with increasing $B$ and $T$, respectively\footnote{
This linear decrease was only observed for the high field range we show.
For lower fields, $A_{ZBCP}$ decreases nonlinearly with increasing field,
as it is expected for a ZBCP caused by ABS, see the discussion in \cite{Chesca05}.
},
the intersection of the linear fits with the $A_{ZBCP}=0$ axis defines
$B_{ZBCP}$ (with vertical and horizontal error bars, respectively, in
Fig. \ref{fig:4}). This procedure also allows to determine $B_{ZBCP}$
for $B>16\,$T from $A_{ZBCP}(B)_{|T=const}$ (see inset of Fig.
\ref{fig:4}).
$B_{ZBCP}$ increases monotonically with decreasing $T$, extrapolating to $B_{ZBCP}(T = 0) \approx 25\,$T. 
The black and grey lines show $B_{c2}(T)$, obtained from microscopic calculations (with $T_c = 29\,$K and $B_{c2}(0) = 25\,$T), assuming either a 2D Fermi cylinder (black line) \cite{Dahm03} or a 3D Fermi sphere (grey line) \cite{Helfand66}. These lines may be considered as a lower bound for the true $B_{c2}$ of our samples. Note that $B_{c2}(0)=25\,$T is at least a factor of 3.5 larger than the zero temperature extrapolation of $B_{90}$.

%Figure 4 \begin
%Figure 4 \end

So far we have argued that the observation of the ZBCP extends the superconducting regime to magnetic fields that are much higher than  anticipated. We next should discuss why we see a uniform ZBCP at all at large fields, without evidence for a splitting.  A pronounced splitting would occur due to circulating currents if a single vortex were present near one side of the grain boundary \cite{Graser04}. On the other hand, in the case of a symmetric static vortex configuration, the screening currents that flow along the grain boundary are of equal strengths but opposite direction (cf. arrows in lower right inset in Fig. \ref{fig:2}b). In this situation the shifts of the ABS (forming fast compared to the time scale of vortex fluctuations) should cancel and no splitting is expected. In the case of a vortex liquid, which is likely to be present in the high field region we discuss here, the screening currents on the two sides of the GBJ will fluctuate, but are likely to have zero average. Thus again the energy shifts of the ABS will be washed out during the time of measurement, naturally explaining the experimental observation. On the other hand, at lower fields, where static but disordered vortex lattices may form, some net screening current is flowing along the GBJ. In this case the ABS at zero energy should undergo a Doppler splitting \cite{Graser04}.
%and the ABS should undergo a shift away from the excitation threshold.
Indeed we observe that the ZBCP becomes structured and often asymmetric in fields below the irreversibility line, possibly allowing to use the ZBCP to probe details of the vortex lattices. A detailed discussion, however, is beyond the scope of this paper and will be presented somewhere else \cite{Wagenknecht07b}.

Returning to the QTS as shown in Fig. \ref{fig:2}, we note that the coherence peaks -- often used as an indicator for the superconducting state -- are suppressed already at 50 per cent of $B_{c2}$ or so. This observation is indeed consistent with calculations based on the quasiclassical Eilenberger equations of the density of states in the vortex state, as given in Fig. 5 of Ref. \cite{Dahm02} where, at 0.5$\,B_{c2}$ the coherence peaks are hardly visible. 

In conclusion, we have shown that a zero bias conductance peak can be seen in the quasiparticle tunneling spectra of LCCO grain boundary junctions at magnetic fields that are far above previous estimates of $B_{c2}$. Particularly, a zero temperature extrapolation suggests that $B_{c2} (0)$ is at least 25 T. Extending superconductivity to such high fields shrinks the region where a pseudogap phase may exist. 

This work was supported by the Deutsche Forschungsgemeinschaft(DFG) through projects Kl930/11-1 and DA 514/2-1.
M. W. gratefully acknowledges financial support by the Friedrich-Ebert-Stiftung, Bonn.

%
%bib section starts here
\bibliography{ZBCA-Hc2}
%end of bib section
%
\end{document}